\PassOptionsToPackage{numbers}{natbib}
\documentclass{article}

\usepackage[final]{neurips_2021}

\usepackage[utf8]{inputenc} 
\usepackage[T1]{fontenc}    
\usepackage{hyperref}       
\usepackage{url}            
\usepackage{booktabs}       
\usepackage{amsfonts}       
\usepackage{nicefrac}       
\usepackage{microtype}      
\usepackage{xcolor}         
\usepackage{graphicx}
\usepackage{subfig}

\providecommand{\myceil}[1]{\left \lceil #1 \right \rceil }

\title{Compressing (Multidimensional) Learned Bloom Filters}

\author{%
  Angjela Davitkova \\
  TU Kaiserslautern (TUK)\\
  Kaiserslautern, Germany \\
  \texttt{davitkova@cs.uni-kl.de} \\
  \And
  Damjan Gjurovski \\
  TU Kaiserslautern (TUK) \\
  Kaiserslautern, Germany \\
  \texttt{gjurovski@cs.uni-kl.de} \\
  \And
  Sebastian Michel \\
  TU Kaiserslautern (TUK) \\
  Kaiserslautern, Germany \\
  \texttt{michel@cs.uni-kl.de} \\
}

\begin{document}

\maketitle

\begin{abstract}
Bloom filters are widely used  data structures that compactly represent sets of elements.
Querying a Bloom filter reveals if an element is not included in the underlying set or is included with a certain error rate.
This membership testing can be modeled as a binary classification problem and solved through deep learning models,
leading to what is called learned Bloom filters.
We have identified that the benefits of learned Bloom filters are apparent only when considering a vast amount of data, and even then, there is a possibility to further reduce their memory consumption. For that reason, we introduce a lossless input compression technique that improves the memory consumption of the learned model while preserving a comparable model accuracy. 
We evaluate our approach and show significant memory consumption improvements over learned Bloom filters. 
\end{abstract}

\section{Introduction}
\label{sec:introduction}
For decades, relational database systems are the de facto standard for storing enterprise data, 
while only recently there has been a true interest in introducing the benefits of machine learning techniques for relational databases.
Inspired by the premise that each database component can be replaced or enhanced by a counterpart  from the machine learning world, Kraska et al.~\cite{TheCaseForLIS} opened the field of combining machine learning with relational data structures.
For instance, traditional one-dimensional (B-tree or hash index) or multidimensional indexes, can be replaced by regression or classification deep learning models, drastically reducing the space and the time needed for query answering. 

Existence indexes are data structures capable of answering the membership of an element in a given indexed set.
The most prominent existence index is the Bloom filter~\cite{DBLP:journals/cacm/Bloom70}, due to its space efficiency and querying performance.
Previous work suggests the replacement of Bloom filters with deep learning models for classification~\cite{TheCaseForLIS}. 
A further idea~\cite{LiftingTC} suggests a classification model even in the presence of multidimensional data. 
Contrary to Bloom filters that need to index every value combination in the multidimensional data for answering membership queries on subsets, learned multidimensional Bloom filters can easily infer such pattern connections. 
However, unlike other traditional indexes which in most scenarios are  slower and much larger than their learned parallel, it is difficult to replace Bloom filters, as they are already extremely compact and fast. 
To exhibit a real impact by their replacement, the data has to be  large, typically ranging in billions of records. Intuitively in the cases of multidimensional data, the number of combinations needed to be indexed by the Bloom filter often creates such data. Still, the number of distinct values per dimension also increases the parameters of the neural network and the embedding matrices, affecting the space benefits of the learned model.

In this work, we suggest the usage of a compressed learned Bloom filter which exhibits the benefits of the learned multidimensional existence index while providing drastic space reduction, faster training time, and comparable accuracy. We employ lossless compression of the categorical data which offers benefits over Bloom filters even when the underlying data is small.

\section{Related work and background}
\label{sec:related-work}

\subsection{Related work}

Kraska et al.~\cite{TheCaseForLIS} first suggested the idea of replacing indexes with deep learning models.
Since this premise, a plethora of papers focus on improving traditional one-dimensional~\cite{DBLP:conf/sigmod/DingMYWDLZCGKLK20,DBLP:journals/pvldb/FerraginaV20,DBLP:conf/sigmod/KipfMRSKK020} and also multidimensional indexes~\cite{TheMLIndex,Tsunami} with learned models. 
Since the idea of replacing a traditional Bloom filter~\cite{DBLP:journals/cacm/Bloom70} with a classification model~\cite{TheCaseForLIS}, several improvements have been proposed. For instance, Vaidya et al.~\cite{PLBF} use multiple more accurate learned filters based on classification score segments. Mitzenmacher~\cite{SandwichedLBF} proposes a new sandwiched learned Bloom filter, including two surrounding Bloom filters for improved performance. 
Differently, Macke et al.~\cite{LiftingTC} focus on extending the learned Bloom filter and show the benefits of having such an index for multidimensional data. 
Compressing the input of deep learning models has also been investigated for  learned cardinality estimators. NeuroCard~\cite{DBLP:journals/pvldb/YangKLLDCS20} uses a variable byte compression of columns to improve the proposed cardinality estimator. 
In our experiments, we compare only with the multidimensional learned filter since ideas like partitioning or sandwiching are orthogonal and can be used in combination with our approach.

\subsection{Background}
\label{ssec:background}
\textbf{Bloom filters} are space-efficient data structures, used to test the existence of an element in a given dataset.
Their probabilistic nature enforces the guarantees of no false negatives and tunable false positive rates.
More specifically, a Bloom filter is an $m$ bit array (initially all bits are set to $0$), requiring $h$ well-defined hash functions. Upon adding an element, each of the  $h$ hash functions maps the element to a position in the $m$ bit array and sets the bit to $1$.  
Evidently, the adaptation of a Bloom filter for multidimensional data would require  indexing of {\it all possible} combinations of column values, to be able to accurately decide on the presence of subsets of values.

\noindent
\textbf{Learned Bloom filters} exploit the idea that classification tasks resemble the behavior of Bloom filters. The classification model learns to identify the presence of the elements in the set, by learning on positive samples drawn from the given dataset and negative samples, representing data not present in the given set. The learned model has a smaller memory consumption at the price of increased false positive rate and the introduction of false negatives. To solve the problem of false negatives, previous work~\cite{TheCaseForLIS,LiftingTC} suggests the use of a backup \textit{fixup filter} that stores the false negatives. 
The benefits of the learned Bloom filter are evident for larger dimensions because unlike the traditional Bloom filter which needs to contain all combinations of column-value pairs, the learned model can infer these interconnections.
The multidimensional learned Bloom filter~\cite{LiftingTC} considers $n$ string tuples, each first converted into an embedding vector. 
The embedding vectors are concatenated and fed through dense layer(s). Using the sigmoid activation, the output is converted to a logit suitable for presenting the presence and absence of terms. We extend on this idea, by improving space consumption.

\section{Compression for learned Bloom filters}
\label{sec:compression-for-learned-bloom-filter}

\subsection{Modeling multidimensional data}
As an example for multidimensional data, consider a dataset containing vehicle information for a car rental agency. 
For simplicity, let us consider that the dataset has only three columns, the name of the car, the fuel type, and information on whether the car has been rented out. 
Intuitively, indexing multidimensional data would lead to a more comprehensive Bloom filter since the filter needs to answer membership queries involving subsets of the original records. 
As an example, considering the vehicles dataset described above, checking if there is an available car that runs on diesel can be done by querying for \textit{(${}?$,diesel,true)} where "$?$" is a placeholder for any value. To answer this query, the Bloom filter needs to index all subsets of column values for a particular entry, including the ones where the column values are not specified, i.e., queries of the type \textit{($?$,fuel\_type,rented\_out)}.
Thus, the size of the Bloom filter will be directly affected by the possible combinations of co-occurring attributes that need
to be covered by the index. On the other hand, when considering learned Bloom filters over multidimensional data, the model parameters are directly affected by the input dimensions, i.e., the distinct values of the columns. Hence, datasets in which columns have many distinctive values can result in higher memory consumption of the learned Bloom filter, although the space consumption will be still lower compared to traditional Bloom filter implementations~\cite{TheCaseForLIS,LiftingTC}. 

\begin{figure}[!t]
\vspace*{-2mm}
  \center
  \includegraphics[width=0.9\linewidth]{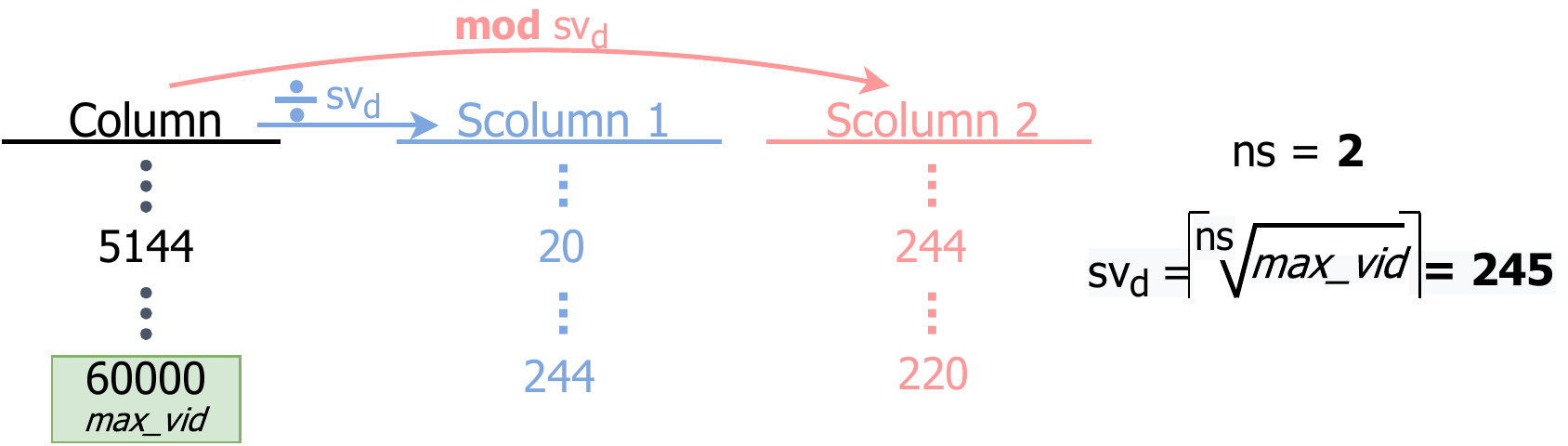}
  \caption{Compressing a column into two subcolumns}
  \label{fig:compression_example} 
\end{figure}

\subsection{Lossless input compression}

The number of inputs and their distinct values have a direct impact on the model parameters and, thus, the size of the model. One way to provide the inputs to the model is to use an embedding layer on every column. To accomplish this, proper mapping of string data to an integer value is performed for every column. However, the size of the embedding matrix will scale linearly with the number of unique values per column. Thus, even for a column with $10^5$ unique values, when using a $32$-dimensional embedding, the embedding matrix would take around $12.8$MB of space which is already much larger than a normal Bloom filter. By applying our proposed input compression, we aim at drastically reducing this space consumption.   

Consider a relation $R$ where $c_1$, $\dots$, $c_n$ represent columns with $v(c_1)$, $\dots$, $v(c_n)$ unique values.  
The main idea is to split a column into several subcolumns, together having fewer dimensions than the original column, contributing to a smaller encoding of the input and, thus, a smaller model size. 
The number of subcolumns is chosen based on the number of distinct column values, with the goal of smaller input dimensionality. 
For example, if the number of values for column $c_i$ is $v(c_i)=10000$, two subcolumns are sufficient to efficiently compress the original column values whereas for a column with $10$ million unique values three or more subcolumns would be required to train the model. Our compression is based on the observation that we can reduce the input dimensionality by dividing the column values with a specific divisor. 
More specifically, to apply the encoding, we first identify the number of subcolumns $ns$ that a column should be split into. Then, we set as a divisor $sv_d$ to be the $ns^{th}$ root of the number of distinct values of the column, i.e., $sv_d\!=\myceil{\!\sqrt[ns]{max\_vid \ } \ }$. For compressing the column value $x$, we determine the quotient $sv_q$ and reminder $sv_r$ when dividing the value $x$ with $sv_d$. If $ns > 2$, we repeat the same procedure for $x = sv_q$ and $max\_vid = max\_sv_q$, at the end reaching $ns$ subcolumns. 

As an example, consider the column represented in \figurename~\ref{fig:compression_example}. As depicted, the number of distinct values for the column is $max\_vid = 60000$ and we want to compress the value $x = 5144$. 
If we want to split the column into two subcolumns, i.e., $ns = 2$, then as the divisor we get $sv_d = 245$ by calculating the squared root of $max\_vid$. 
Thus, the value $x = 5144$ will be compressed in $sv_q = 20$ and $sv_r = 244$. In this case, all column values are split into two subvalues having maximal values $sv_d$ and $sv_d - 1$. Through the compression, we reduce the number of dimensions needed to encode the input from $60000$ to $489$. When considering a $32$-dimensional embedding, we reduce the size of the embedding matrix from $7.8$MB to approximately $0.06$MB, which is a substantial space reduction.  

We perform the compression over every column for which $v(c_i)$ is greater than a compression threshold $\theta$. Unlike previous learned multidimensional Bloom filters, which encode the structure of the column through embeddings, we also allow a one-hot encoding in cases where the column has been already compressed to a smaller dimension and embedding matrices are no longer necessary.

Evidently, the proposed input compression would affect not only the model size but also the model accuracy. More specifically,  parameter $ns$,  which determines the number of subcolumns a column needs to be split into, allows a tradeoff between the model size and model accuracy. By increasing $ns$, although the unique column values are decreased, the number of input columns is increased. Consequently, the learning of the model will be negatively affected since it would need to learn across multiple columns with increased interconnection. Thus, we carefully set both $ns$ and the compression threshold $\theta$ for achieving an acceptable tradeoff.

\section{Experiments}
\label{sec:experiments}

\noindent
\textbf{Setup\&Datasets:}
For generating positive training data, we randomly sample from the  data records and optionally replace some of the values with wildcards.
For negative training data, we randomly select non co-occurring combinations of values,  optionally including a wildcard.
If not mentioned otherwise, columns are compressed into $2$ subcolumns. 
We implemented our proposed model in Keras, Python, and  
performed the experiments on NVidia GeForce RTX $2080$ Ti GPU.
The experiments were performed using two real-world datasets, where we retrieve $100,000$ records, following the analysis of previous work~\cite{LiftingTC}.
The first dataset (airplane) consists of flight information, with $7$ columns having $v(c_i) = [6887, 8021, 8046, 6537, 2557, 5017, 1663]$.
The second dataset (DMV) consists of vehicle registration data~\cite{DmvData} which has $19$ columns, having less distinct values than the airplane dataset, i.e., $v(c_i) = [5, 10001, 27, 1627, 27, 1570, 64, 107, 694, 40, 8, 1509, 346, 966, 794, 102, 3, 3, 2]$. 
The different distributions outline the benefits and drawbacks of 
our compression approach 
 ({\bf C-LMBF}).  We compare with the original learned multidimensional filter ({\bf LMBF}) and traditional Bloom filter ({\bf BF}). For BF, we only use $\approx5$ million unique subset combinations. 
 Since the accuracy and FNR of C-LMBF and LMBF are almost identical, we do not show the fixup filter memory in the experiments.

\begin{figure}[!t]
\vspace*{-4mm}
\centering
\begin{minipage}[!t]{0.9\textwidth}

  \begin{minipage}[!t]{0.5\textwidth}
  \captionof{table}{\label{tab:comparison} Comparison of C-LMBF, LMBF (both with 1 layer of 64 neurons) and BF }
    \resizebox{\columnwidth}{!}{%
    \centering

  \begin{tabular}{|l|c|c|c|c|}
    \hline
        \multicolumn{5}{|c|}{Airplane} \\ 
            \hline
    \multicolumn{1}{|c}{} &
      \multicolumn{1}{c|}{} &
      \multicolumn{3}{c|}{Memory} \\
      \hline
     & Accuracy &  MB & NN params & Input dim \\
    \hline
    $\theta$ = 3000 & 0.95 & 0.53 & 33,006 & 5060 \\
    \hline
    $\theta$ = 5500 & 0.97 & 1.01 & 73,110 & 9933 \\
    \hline
    $\theta$ = 8000 & 0.98 & 2.35 & 186,713 & 23025 \\
    \hline
    LMBF & 0.98 & 4.06 & 330,608 & 38728 \\
    \hline
    BF-0.1 & 1 & 6.10 & --- & --- \\
    \hline
     \hline
            \multicolumn{5}{|c|}{DMV} \\
            \hline
     \multicolumn{1}{|c}{} &
      \multicolumn{1}{c|}{}&
      \multicolumn{3}{c|}{Memory} \\
      \hline
     & Accuracy & MB & NN params & Input dim \\
    \hline
    $\theta$ = 100 & 0.98 & 0.36 & 5,447 & 892 \\
    \hline
    $\theta$ = 1000 & 0.98 & 0.47 & 19,564 & 3636 \\
    \hline
    $\theta$ = 2000 & 0.98 & 0.78 & 47,694 & 8097  \\
    \hline
    LMBF & 0.98 & 1.97 & 147,351 & 17895  \\
    \hline
    BF-0.1 & 1 & 6.10 & --- & ---  \\
    \hline
  \end{tabular}
  }
      
  \end{minipage}
   \hfill
  \begin{minipage}[!t]{0.45\textwidth}
      \vspace*{8.5mm}
      \centering
      \includegraphics[width=\linewidth]{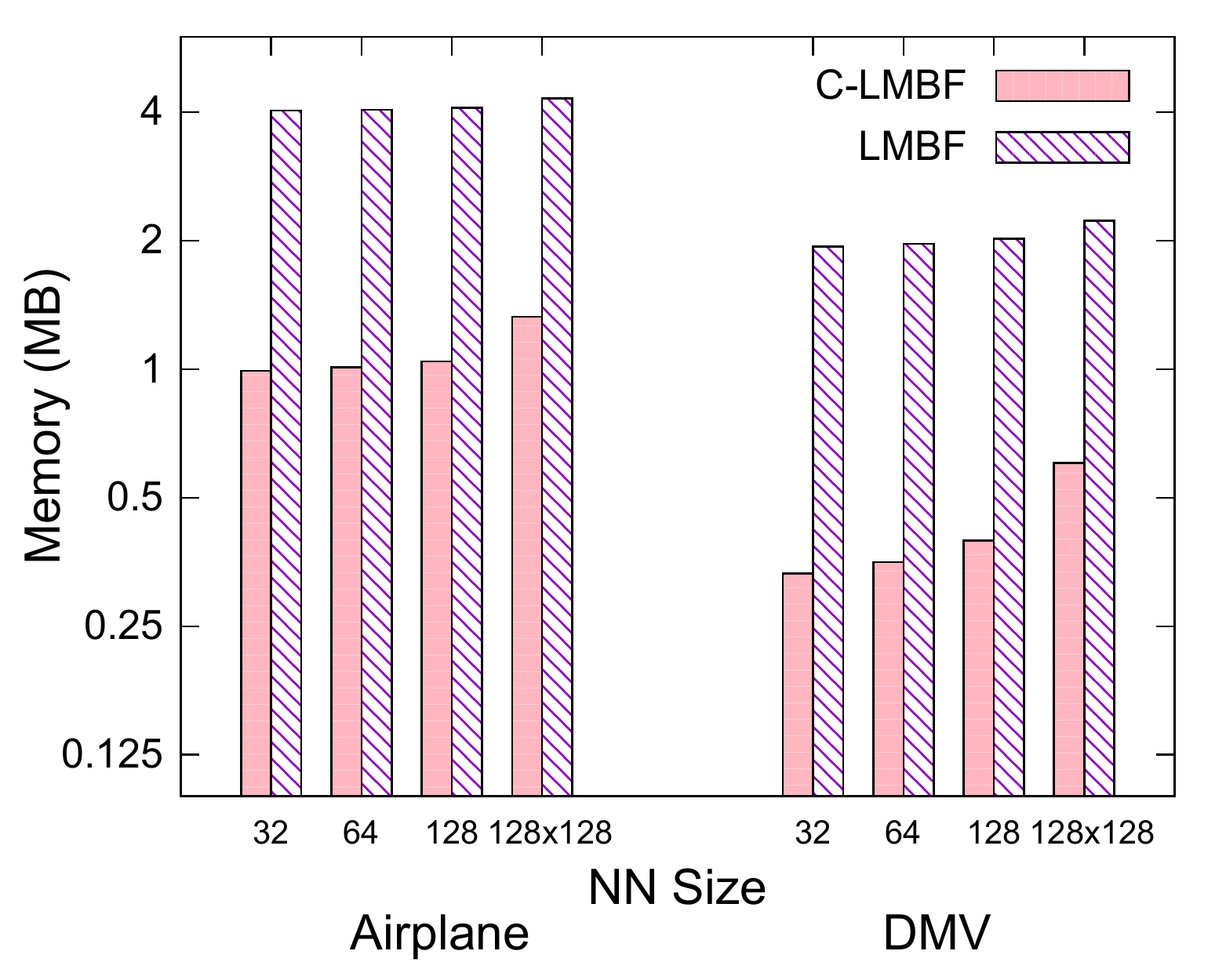}
      \captionof{figure}{Memory when varying NN size}
      \label{fig:spec-vs-comb-star}
  \end{minipage}
  
\end{minipage}
\end{figure}

\noindent
\textbf{Results:}
In Table~\ref{tab:comparison}, we show several measurements for C-LMBF with different compression boundaries $\theta$. 
The different $\theta$ boundries result into $[5, 4, 2]$ and $[10, 4, 1]$ compressed columns, for each dataset respectively. 
We fix the number of layers and dimensions for each of the models and train them until convergence. 
The embedding is set according to the input dimension size. 
The compression reduces up to $\approx7$x and $\approx20$x of the input dimensionality, for the airplane and DMV dataset respectively, with a trade-off of a small accuracy reduction. The decrease in the input dimensions shows a drastic reduction of the memory of the model as well as the needed parameters for training. Since the columns with the smaller number of values do not require compression, it is not unusual that a smaller $\theta$ may introduce additional complexity, and thus produce worse accuracy, as most evident for the airplane dataset when $\theta = 3000$. 
In Figure  \ref{fig:spec-vs-comb-star}, we show the impact of  different neural network sizes on the memory consumption. We set  $\theta = 5500$ for the  airplane dataset and  $\theta = 100$ for the DMV dataset. As expected, the C-LMBF has a constant reduction in size when compared to LMBF. Furthermore, the increase in the neural network size causes a better or an equal accuracy (not shown). 
The number of subcolumns created ($ns$) also impacts the model.
The considered datasets have fewer unique values per column and dividing into subcolumns for $ns>2$ would only increase the number of inputs and embedding matrices without a beneficial reduction on the input dimensions. Although not shown, larger values of $ns$ are highly useful for many distinct values, e.g., considering knowledge graph data.
For a carefully chosen $\theta$ and $ns$, the compression also causes faster training time, e.g., we have a $10$--$15$ seconds speedup when executing $1$ epoch.

\section{Conclusion}
\label{sec:conclusion}

\vspace{-1mm}

We  introduced a memory efficient learned Bloom filter by compressing the input parameters of the model using a lossless column compression that splits the values of the input columns into a predefined number of subcolumns. Through  experiments, we showed drastic memory consumption improvements while keeping comparable accuracy.

\bibliographystyle{plain}
\bibliography{neurips_2021}

\end{document}